# Compact multi-channel radio frequency pulse sequence generator with fast switching capability for cold atom interferometers

Min Jiang,[1] Si-Bin Lu*,[1] Yang Li,[1,2] Chuan Sun,[1,2] Zhan-Wei Yao,[1,3] Shao-Kang Li,[1] Hong-Hui Chen,[1,2]
Xiao-Li Chen,[1,2] Ze-Xi Lu,[1,2] Yin-Fei Mao,[1,2] Run-Bing Li*,[1,3,4] Jin Wang,[1,3,4] and Ming-Sheng Zhan[1,3,4]

[1]*State Key Laboratory of Magnetic Resonance and Atomic and Molecular Physics,*
*Innovation Academy for Precision Measurement Science and Technology, Chinese Academy of Sciences, Wuhan 430071, China*
[2]*School of Physics, University of Chinese Academy of Sciences, Beijing 100049, China*
[3]*Hefei National Laboratory, Hefei 230088, China*
[4]*Wuhan Institute of Quantum Technology, Wuhan 430206, China*
(*rbli@wipm.ac.cn)
(*lusibin@apm.ac.cn)
(Dated: March 1, 2023)

Cold atom interferometers have matured to a powerful tool in fundamental physics research, and they are currently on their way from realizations in the laboratory to applications in the real world. The radio frequency (RF) generator is an indispensable component to control lasers and then manipulate atoms. We developed a highly compact RF generator for fast switching and sweeping frequencies/amplitudes of atomic interference pulse sequences. Multi-channel RF signals are generated by using a field-programmable gate array (FPGA) to control eight direct digital synthesizers (DDSs). We further proposed and demonstrated a method of preloading the parameters of all RF pulse sequences to the DDS registers before the execution of the pulse sequences, which eliminates the data transfer between the FPGA and DDSs to change RF signals and thus sharply shortens the delay of frequency switching when the pulse sequences are running. The characterized performance shows the generated RF signals achieve 119 ns frequency switching delay, and 40 dB harmonic rejection ratio. The generated RF pulse sequences are applied to a cold atom-interferometer gyroscope, and the contrast of atomic interference fringes reaches 38%. This compact multi-channel generator with the fast frequency/amplitude switching or/and sweeping capability has beneficial applications for the real-world atom interferometers.

## I. INTRODUCTION

Cold atom interferometers (CAIs), including atomic gyroscopes [1–3], gravimeters [4–8], and gravity gradiometers [9–11], have wide applications in precision measurement fields. With CAIs, the fine structure constant $\alpha$ [12] and gravitational constant $G$ [13] were measured, and the Einstein's equivalence principle was also tested [14]. Moreover, the gravitational wave can be detected [15], and the gravity-field curvature be measured [16] with high-precision CAIs. In the atom interferometer, atoms are manipulated by a series of complex light pulse sequences which are generated by acousto-optic modulators (AOMs) controlled by radio frequency (RF) signals, i.e., Fig. 1 (b) shows the pulse sequences needed by an atom-interferometer gyroscope, where their pulse width should be as short as possible. As a consequence, a multi-channel RF pulse sequence generator with fast switching capability is indispensable for CAIs. Conventionally, RF pulse sequences were generated by commercial generators along with RF switches [17]. These devices take up large spaces and consume high powers, which frustrates the real-world applications required by the low Size, Weight and Power (SWaP) [18]. The RF pulse sequence can also be generated by a voltage-controlled oscillator, but it has drawbacks of the low frequency resolution, stability, and modulation rate [19].

On the contrary, RF signals produced by the digital direct frequency synthesizer (DDS) has high stability, modulation rate, fine resolution, and wide frequency adjustment range [20]. With field-programmable gate arrays (FPGA) and DDSs, the single-channel [21–24] and multi-channel [20, 25–27] RF signal generators were developed for experiments in cold atomic physics. In above cases, user input parameters are received by the top-level FPGA, then transferred to next-level multiple FPGAs, and eventually loaded into the DDS chips by this last-level FPGAs. Because control parameters are transferred several times and finally loaded into the DDS register for changing the frequency and/or amplitude of DDS output signal during the generation of a RF pulse sequence, there is a delay time of a few tens of microseconds forcing the new frequency and/or amplitude to appear at the RF generator outputs. To our knowledge, there are few studies that can produce all the RF pulse sequences with the capability of fast switching frequency and/or amplitude less than a few microseconds, required by atom interferometers. On the other hand, those generators are not compact enough to generate multi-channel RF signals, needed by the real-world applications of CAIs.

In this paper, we present a compact, multi-channel RF pulse sequence generator with fast switching capability for CAIs by exploiting an FPGA to control multiple DDSs. The control parameters of all the RF pulse sequences needed by CAIs are loaded into the DDSs register before the starting of the pulse sequence, which eliminates data transfer between the FPGA and the DDS chips when the pulse sequence is running, thus sharply shortening the switching time of frequency and/or amplitude to 119 ns during pulse sequence execution. Benefit from the parallel calculation capability of FPGAs, the synchronism of generated RF pulse sequences can be ensured. By placing the components on the printed circuit board (PCB) whose size is only $15 \times 15$ cm$^2$, and by carefully routing the tracks on multi layers, the phase noise of the generated RF signal is achieved to $-72$ dBc/Hz@1Hz when the carrier frequency is 98.6 MHz, and the harmonic rejec-tion ratio is reached to 40 dB when the carrier frequency is up to 400 MHz. This multi-channel RF pulse sequence generator was applied to a dual-loop atom-interferometer gyroscope and atomic interference fringes were observed with the contrasts of 38%. In Sec. II, we will describe in detail the proposed scheme including hardware and software design. In Sec. III, the performance of the RF generator is characterized. Finally, a summary is given and applications are discussed in Sec. IV.

## II. DESIGN CONCEPT

### A. Overview

In CAIs, different kinds of light pulse sequences are needed to manipulate atoms, as shown in Fig. 1 (b). First, two-dimensional cooling lights are utilized to pre-cool the atoms and three-dimensional cooling lights are applied to trap the atoms in a Magneto-Optical Trap (MOT), where repumping lights are also needed. In order to cool atoms to sub-Doppler temperature, the polarization gradient cooling (PGC) is usually applied, which requires to fast sweep the frequency and intensity of cooling lasers synchronously. Second, Raman and blow-away lights are used to selects atomic velocities and prepare atomic initial states, and Raman lights are also used to coherently manipulate atomic wave packets with a timing sequence of three pulses ($\pi/2$-$\pi$-$\pi/2$). Finally, to observe atomic interference fringes, the population of coherent states is detected by probe lights which are provided by switching and shifting the frequencies of cooling and repumping lights. Therefore, the fast switched and swept light pulse sequences are highly required for building the CAIs, which are usually realized with the AOMs driven by the RF pulse sequences.

To generate above required RF pulse sequences, we employ an FPGA to control eight DDSs, as shown in Fig. 1 (a). The F-PGA is the core module to control the timing sequences of the eight-channel RF pulses, which connects to a user-interface computer via a USB connector for data transfer and to the DDS chips with serial I/O, PROFILE, and DRG pins. Once the rising edge of an external trigger signal is detected by the FPGA, eight-channel RF pulse sequences are parallelly generated. At the same time, an external input clock signal is distributed to the FPGA and DDS chips with a clock buffer chip as their clock sources, and a 12 V power rail is required to power the FPGA and DDSs after being converted to lower voltages with 5 V and 3.3 V DC/DC converters and low dropout regulators (LDOs). Making full use of DDS chip operation modes, we generate the multi-channel RF pulse sequences with fast switching capability for CAIs, required by Fig. 1 (b). In the single-tone mode, the frequency of a RF signal can be changed at most eight times, which is sufficient for controlling the RF pulse sequences required by the two-dimensional cooling light, Ranman light, blow-away light and repumping light. For the complicated pulse sequence, such as the three-dimensional cooling light, the frequency and amplitude of the RF signal need to be synchronously controlled. With the RAM mode, the whole waveform of the RF pulse can be split into eight segments of frequencies, and each of them can initiate a new waveform when it is selected by PRO-FILE pins. The frequency of the RF signal is controlled by the RAM mode and its amplitude swept by the DRG mode at the PGC stage. Thus, the whole RF pulse sequences are generated by designing the FPGA software with these three modes.

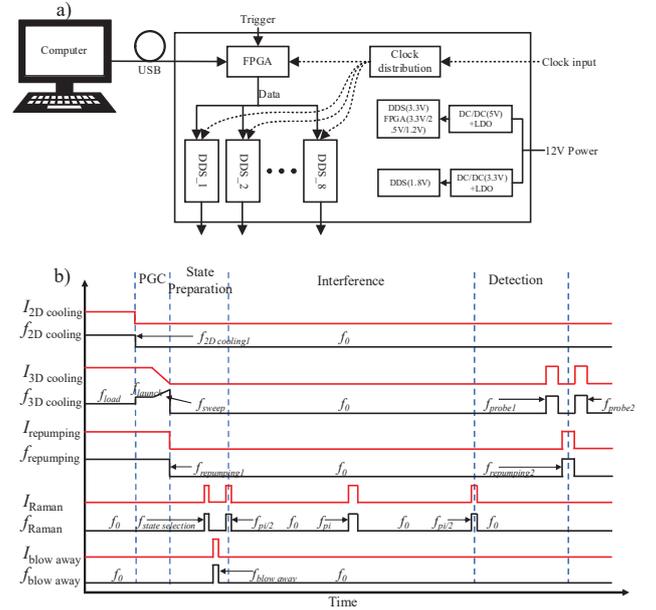

FIG. 1: Block diagram of the multi-channel RF generator (a) and RF pulse sequences for a cold atomic gyroscope (b).

### B. Hardware design

The atomic interference process requires RF signals can be switched and swept according to a temporal schedule with a strict resolution of less than a few microseconds. The relative delay of starting times among different RF pulse sequences should be as short as possible. Compared to digital signal processors or single-chip computers which execute commands sequentially, the FPGA chip is an efficient mean to reliably and precisely control the multi-channel RF signals since it has the powerful capability of parallel calculus and ensures the synchrony of signals. Here, an FPGA chip (EP4CE6E22C8) from Altera Corporation is chosen as the control unit, and eight DDS chips (AD9910) from ADI Corporation are employed to generate multi-channel RF signals. This DDS chip with a maximum clock frequency of 1 GHz provides a RF signal up to 400 MHz with a resolution of 0.23 Hz, and integrates an internal volatile memory, where a table of arbitrary frequencies is stored and later recalled automatically with a programmable delay between consecutive steps [28].

The component layout is crucial to generate an RF signal with good performance. Digital and analog signals should be routed as far as possible to eliminate the interference between two kinds of signals. As shown in Fig. 2, the FPGA, tracks connecting the FPGA and the DDSs are placed in the digital sector which is in the central area of the PCB in our layout.

The analog signals are routed in two sides of the digital sector. The complementary output current signals of the DDS are transformed into the single-end voltage signal by the Balun transformer (ADT1-AWT). After analog voltage signals are further low passed by filters (LFCN-400) and amplified by fixed gain amplifiers (MAR-4SM+) to about 5 dBm, they are outputted as RF signals from both sides of the PCB.

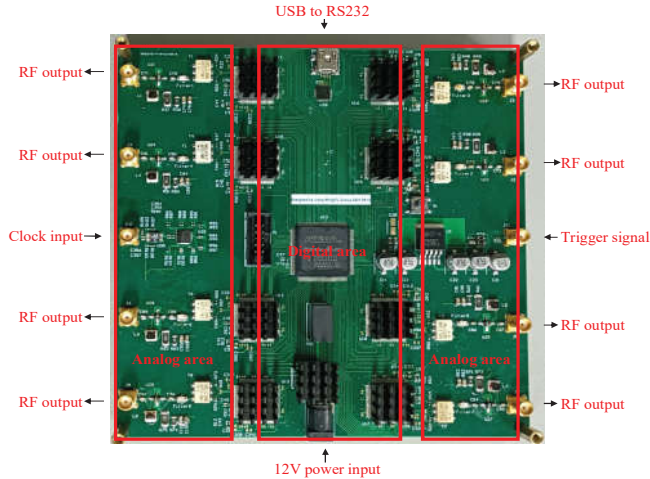

FIG. 2: PCB of the FPGA and DDS board with main input/output ports and separated digital/anolog areas labeled.

On the left side of the PCB, there is a clock signal input port apart from four RF signal output ports. Firstly, the single-end input clock signal is transformed into a low voltage differential signaling (LVDS) signal by an AC signal logic level translator (LT6957-2) to increase the common-mode noise suppression. After feeding the obtained LVDS signal into the clock distributor (SI53344), nine LVDS clock signals are provided to the FPGA and eight DDSs, respectively. To avoid the energy reflection between the clock signal transducer and receiver, the differential impedance of the differential signal track is matched to 100 Ohms. On the right side of the PCB, there is an external trigger signal input port apart from another four RF signal output ports. By employing a USB to RS232 chip (CP2102N) which is located in the upper side of the PCB board, FPGA can load the user input parameters, such as RF signal frequency or/and amplitude, into the internal register of the AD9910 via the serial I/O port. The PCB operates from a 12 V supply and has on-board DC/DC modules and LDOs to generate the required power rails for the onboard components. This combination can improve the efficiency of power conversion and the capability of power ripple rejection. To eliminate mutual interference between digital chips and analog chips, they are powered by separate regulators. AD9910 includes many function modules, such as serial communication module, PLL module, and DAC module, etc. To achieve the highest performance, those modules should be powered separately. For example, the digital I/O port, DAC module, and PLL module should be respectively powered by 3.3 V DVDD, 3.3 V AVDD and, 1.8 V AVDD. Eventually, the power of designed PCB typically consumes about 18 W at full operation, and the area of the designed PCB occupies $15 \times 15$ cm$^2$, which are far more compact than commercial RF generators.

## C. Software design

To generate the light pulse sequences, the FPGA software needs to be carefully designed. In Fig. 1(b), two-dimensional cooling light, Raman light, blow-away light and repumping light are switched by changing the frequency of the RF signal, and the three-dimensional cooling light are controlled by changing both of the frequency and amplitude of the RF signal. It is complicated for designing the pulse sequences of three-dimensional cooling light. Especially, during the polarization gradient cooling stage, we need to linearly increase the frequency detuning and decrease the intensity of the cooling lights. Making full use of eight profile registers in the AD9910 whose addresses range from 0x0E to 0x15, the light pulse sequences mentioned above can be efficiently provided. In the AD9910, different profile registers contain different signal control parameters. Each profile is independently accessible. Using the three external profile pins (PROFILE [2:0]) to select the desired profile register can initiate a new waveform.

Fig. 3 shows the profile register configuration for the generation of Raman light pulse sequence. In this implementation, the single tone mode of the AD9910 is used and profile registers become single tone profile registers, containing the frequency, amplitude, and phase of output signal. RF signal frequencies contained in single tone profiles 4(0x12), 1(0x0F), 2(0x10), and 7(0x15) are used to produce the state selection, $\pi/2$, $\pi$, and $\pi/2$ light pulses, respectively. Selecting single tone profiles 0(0x0E) and 6(0x14) changes RF signal frequency far beyond the modulation bandwidth of AOMs. After configuring the single-tone profile registers, a timer is set for each light pulse on the FPGA. When the timer is timeout, change the states of the PROFILE [2:0] pins to select another set of DDS control parameters contained in the profile register to change the generated RF waveform. In Fig. 3, under the single-tone profile register address, it is the corresponding PROFILE [2:0] pin states. With this kind of single-tone profile register configuration, only one PROFILE pin needs to be changed, when next waveform needs to be initiated. It can avoid the disorder of the produced RF pulse sequence, due to the desynchrony of PROFILE [2:0] pin state change. The single-tone profile register configuration of two-dimensional cooling light, repumping light, and blow-away light is similar to the Raman light and will not be further discussed here.

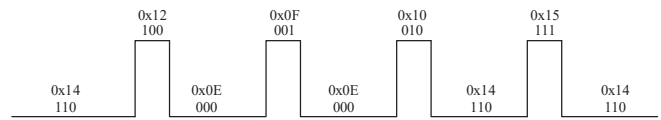

FIG. 3: Configuration of the single-tone profile register for the Raman light pulse sequence.

Figure 4 shows the RAM profile register configuration for the three-dimensional cooling light pulse. To reach sub-Doppler temperature after the PGC stage, the pulse sequence is more complicated because of requirement of successively sweeping frequency detunings and intensities of the three-dimensional cooling light. The frequency and amplitude of the RF signal need to be consecutively swept, which cannot be implemented with the single-tone mode because the output RF signal frequency can be changed eight times at most. Fortunately, taking advantage of the RAM mode, 1024 frequencies can be loaded into the $1024 \times 32$-bit RAM register. The RF signal frequency can be changed at most 1024 times at a specified playback rate. In RAM mode, profile registers become RAM profile registers defining the number of samples and the sample rate for a given waveform. The corresponding number of frequencies are loaded into the RAM register whose address is 0x16. We not only employ the RAM mode to linearly sweep the frequencies of the RF signal but also use the DRG mode to linearly sweep its amplitudes. The RAM profile registers (0x0F and 0x15) are set as the same frequencies and far beyond the modulation bandwidth of AOMs.

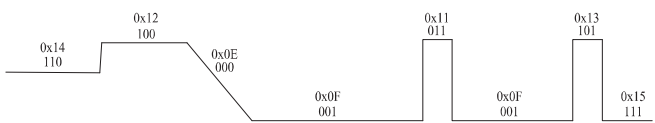

FIG. 4: Configuration of the RAM profile register for the three-dimensional cooling light pulse sequence.

Furthermore, the frequency difference between two Raman lasers needs to be swept to capture the resonant two-photon transition signals, and their phase difference be swept to obtain the interference fringe. In these two scenarios, the frequency and phase parameters should be increased by one step value after the rising edge of the trigger signal being detected. The DRG mode of AD9910 can not only ramp the R-F signal amplitude as mentioned above but also can sweep the frequency or phase of the RF signal. At the same time, the DRG mode also supports a hold feature controlled by the DRHOLD pin. When this pin is set to Logic 1, the DRG is stalled at its last state; otherwise, the DRG operates normally. With these two features, we can increase the RF signal frequency or phase value one step forward per trigger. In our implementation, when the FPGA is triggered by the external signal, the DRCTL pin will be set to Logic 1 to sweep the frequency/phase during the time interval of one more step. Subsequently, the DRHOLD pin will be set to Logic 1 to stall at the desired frequency or phase value for the specified time. Eventually, the DRCTL pin and DRHOLD pin will be set to Logic 0 to ramp with a negative slope in order to restart the ramp from the lower limit after being triggered again.

To conveniently change RF signal parameters, a UART communication protocol is designed. Ten successively received characters are grouped in an eighty-bit packet. The first byte received in each communication specifies the channel number. Following data corresponding to the RF signal control parameters. A state machine on the FPGA interprets the bytes and sends the data to their required location. Using this predefined UART communication protocol, it is usable with a broad range of commercial software programs, such as LabVIEW, or other programming languages. After all the us-er input parameters is received by FPGA, it will reconfigure eight AD9910s to update the output RF pulse sequences.

## III. PERFORMANCES

Leveraging the parallel calculation capability of FPGAs, the eight-channel RF pulse sequences can be synchronously started almost at the same time. All the RF signal control parameters of the temporal sequence of RF pulses are pre-loaded into the internal register of AD9910 and then the sequence is executed after an external trigger is provided. There is no need to reprogram the AD9910 register by the FPGA during the RF pulse sequence generation. Making full use of single tone mode, RAM mode, and DRG mode, only the PROFILE [2:0], DRCTL, DRHOLD pin states need to be changed to initiate a new wave during the sequence execution, thus sharply shortening the delay of frequency switching. This is the main feature of our proposed scheme. We evaluate the performance of the designed RF generator, including spectral purity, phase noise, the delay of frequency switching, the relative delay between starting times of eight-channel RF pulse sequences. These RF pulse sequences are then applied to control AOMs for generating the corresponding light pulse sequences in an actual CAI, and the atom interference fringes are observed.

### A. Spectral purity of multi-channel RF signals

Spectral purity is the inherent stability of an RF signal. Therefore, it is crucial for CAIs to make laser resonate with the atomic transition. With a spectrum analyzer from the Agilent Corporation (N9030A), the parameters of the RF signal are measured, including harmonic and spurious signal, noise floor and phase noise. The spectrum analysis of the output single-tone signal when the carrier frequency is 10 MHz, 98.6 MHz and 397.8 MHz, is illustrated in Fig 5 (a-c). These carrier frequencies are located in the low, median, and high frequency band, respectively. Spectra of eight-channel signals are depicted in different colors and linetypes. In Fig. 5 (a) and (b), the 20 MHz, and 197.2 MHz harmonic signals are respectively corresponding to the 10 MHz, and 98.6 MHz carrier frequencies. In Fig. 5 (c), there are some spurious signals apart from the 397.8 MHz carrier frequency and 795.6 MHz harmonic signal. The harmonic and spurious signals are generated from the internal clock signal of AD9910s when it is converted to the desired carrier frequency. Although eight-channel RF signals are generated at the same time, we did not see any crosstalk among the adjacent DDS channels. In addition, the noise floor power levels are lower than −65 dBm. The harmonic rejection ratio is more than 40 dB and the spectra of eight-channel RF signals have a good consistency.

There are many ways to define spectral purity. The most common and meaningful method of specifying short-term stability is a plot of the signal generators single-sideband phase noise in a 1 Hz bandwidth versus the offset from the carrier. Figure 5 (d), (e), and (f) respectively show the phase noise performance of the eight-channel RF signals with the carri-er frequency being 10 MHz, 98.6 MHz, and 397.8 MHz. As shown in Fig. 5 (e), when the carrier frequency is 98.6 MHz, which is located in the frequency band used by AOM drivers, the phase noise reaches −72 dBc/Hz@ 1 Hz. When the carrier frequencies are 10 MHz and 397.8 MHz, the phase noises achieve −89 dBc/Hz and −56 dBc/Hz at the 1 Hz offset from the carrier. Obviously, the phase noise increases as the carrier frequency increased, while holding the same general shape of the curve. It can be seen from Fig. 5 (d)-(f) that the phase noise performance of the eight-channel RF signals has a good consistency and the low noise floors below −120 dBc/Hz.

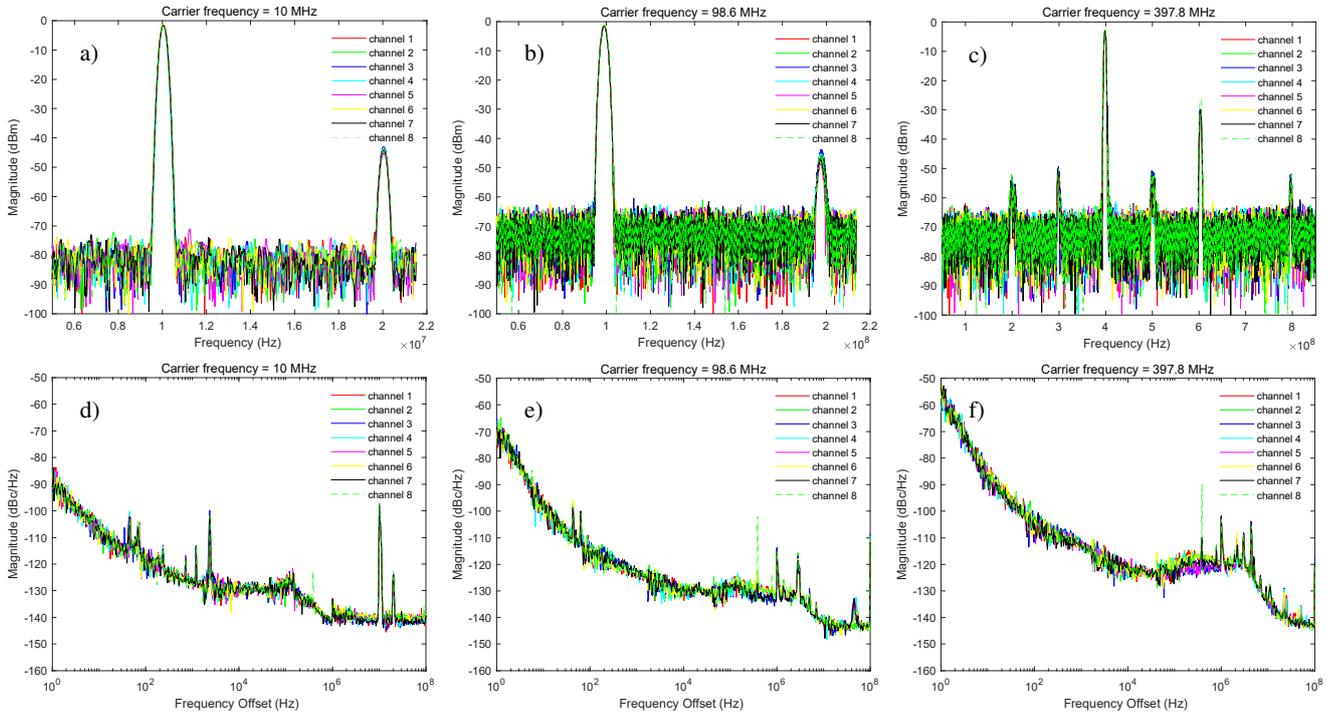

FIG. 5: Spectral purity of generated RF signals. The spectra (a)-(c) and phase noise (d)-(f) among eight-channel RF signals are compared when they are in low, medium and high frequency bands, respectively.

**B. RF pulse sequences with fast switching capability**

Considering the above measured results, the RF signal generated by our designed generator has a good spectral purity performance. To manipulate the atoms, the RF pulse sequences are needed in CAIs. By employing the hardware and software design concept mentioned in Sec. II, the actual eight-channel RF pulse sequences are produced, required by the light pulses as shown in Fig. 1 (b). Figure 6 illustrates RF pulse sequences to generate the three-dimensional cooling, two-dimensional cooling, repumping, Raman, blow away lights. It is worth noting that the values of the RF pulse width and the steps of frequency sweeping in these figures are just for illustration and can be adjusted easily according to the requirement via the USB to UART port. When the rising edge of the externally input trigger signal is detected, these eight-channel RF pulse sequences are synchronously started.

In other words, the FPGA will parallelly switch the frequencies which are beyond the modulation bandwidth of AOMs to the first frequency pulse of eight-channel RF pulse sequences. To manipulate the cold atoms at the given time, the relative delay time among the starting times of the eight-channel RF pulse sequences should be as short as possible. Taking the starting time of one channel RF pulse sequence as the reference, the delay times of other channel RF pulse sequences are measured by the oscilloscope. The result shows that the maximum relative delay time is 3 ns, which is sufficient for CAIs.

After the RF pulse sequences being started, the ability to rapidly switch the amplitude and/or frequency of an RF signal is of paramount importance for CAIs, especially when the light pulse duration is in the time scale of a few microseconds. In our work, the delay of the frequency and/or amplitude changing is constituted of the rising or falling time of the PROFILE [2:0] pins of the DDS chip and the pipeline delay between when the frequency and/or amplitude value changes and when it appears on the output. Due to the fact that control parameters of the RF pulse sequence have been loaded into the internal register of the AD9910 before the execution of the RF pulse sequence, there is no delay for FPGA to reprogram the AD9910 via the serial I/O port after the pulse sequence generation starting. Figure 7 (a) shows the RF signal frequency changes from 30 MHz to 80 MHz. Figure 7 (b) illustrates the amplitude changes from half to full $V_{peak-to-peak}$. The delay time of frequency switching is quantitatively given by measuring the rise time of the light signal. Fig. 7 (c) shows the three-dimensional cooling light pulse sequence generated by the three-dimensional cooling RF pulse sequence as given in Fig. 6, where the upper cross and lower cross respectively refer to the 10% and 90% reference levels. The measured result shows that the frequency switching time is 119 ns. Because the measured result includes the rising time of AOMs, it can give the upper bound of the delay of frequency switching of the RF signal, which is much shorter than $25\mu s$ [20], and $1\mu s$ [27]. It is also worthy noting that the actual light pulse shown in Fig. 7 (c) is exactly as the same shape as the light pulse sequence shown in Fig. 1 (b) required by CAIs.

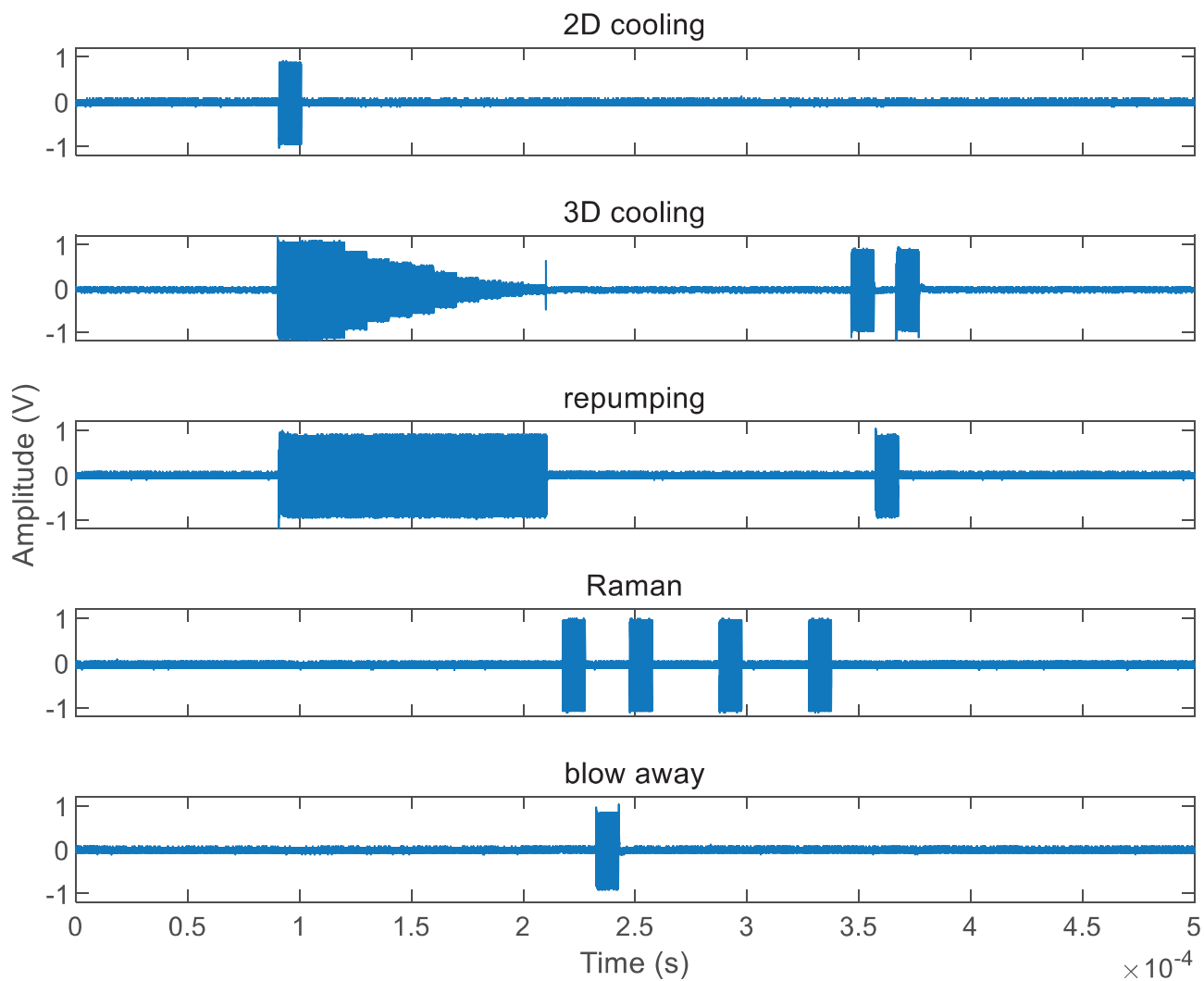

FIG. 6: Switched or/and swept RF pulse sequences for generating the two-dimensional cooling, three-dimensional cooling, repumping, Raman, blow away light pulse sequences in a typical cold atomic gyroscope.

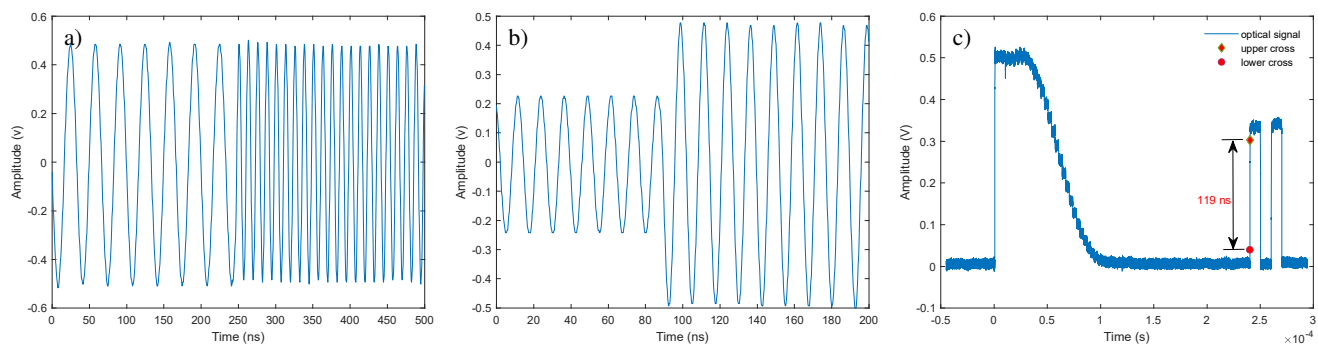

FIG. 7: Waveforms of RF signals when the frequency is switched from 30 MHz to 80 MHz (a), and the amplitude switched from half to full amplitude when the frequency is 80 MHz (b). The light pulse sequence is detected by a fast photo diode (c).

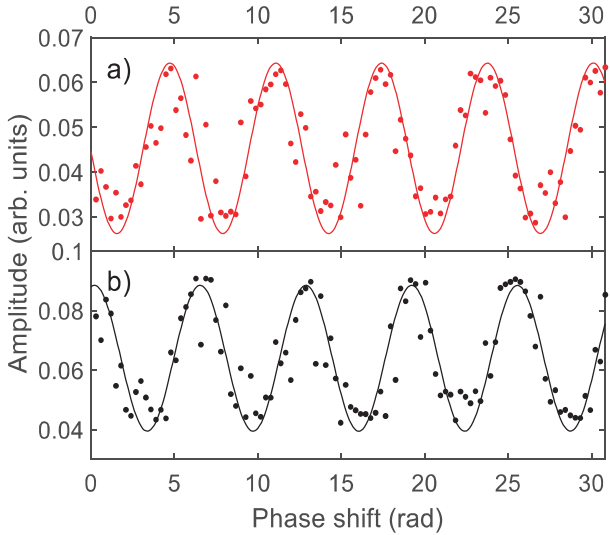

FIG. 8: Mach-Zehnder fringes for the first (a) and second (b) atom interferometers, respectively. Red and black dots are experimental data, and the solid lines are fitted sine curves.

## C. Application to cold atom interferometers

The compact multi-channel RF pulse sequence generator with the fast switching capability is beneficial for the integration of CAIs. The RF pulse sequences generated above are applied to a cold atomic gyroscope recently developed for real-world applications. Similar to our previous work [1], the dual-loop atom interferometer gyroscope is realized by manipulating the atoms with the light pulse sequences, which are generated by driving AOMs with the RF pulse sequences as shown in Fig. 6. As an example, the three-dimensional cooling light pulses shown in Fig. 7 (c) are used to cool and detect the atoms by controlling the frequency and intensity of the lasers. After the atoms are two-dimensionally pr-cooled, they are loaded into two MOTs, and then launched by the moving optical molasses technique. At the same time, the atoms are further cooled by the PGC method, and propagate along the symmetrical parabolic trajectories with opposite directions in the same vacuum chamber. The atomic velocity is 4 m/s, and its temperature is 8 $\mu$K which is further cooled to 0.8 $\mu$K by using the velocity-sensitive Raman transitions. Three pairs of separated Raman lights with an interrogation arm of 12 cm are counter-propagating along the gravity direction. After the atoms are prepared to an initial state with the Raman and blow away light pulses, three Raman light pulses ($\pi/2$-$\pi$-$\pi/2$) are applied to build dual Mach-Zehnder atom interferometers, where the duration of the $\pi/2$ light pulse is 8 $\mu$s. Considering the uncertainty of timekeeping, the delay time of frequency switching of 119 ns is sufficient to generate this Raman light pulse. When the interrogation time is 30 ms, the atomic interference fringes are observed, as shown in Fig. 8 (a) and (b), and their contrasts are reached to 38%. Thus, by using our designed generator, a dual-loop atom-interferometer gyroscope is realized with each of the interference-loop area of 43 mm².

## IV. CONCLUSION

We designed and realized a highly compact multi-channel RF signal generator based on FPGA and DDS. Taking advantage of FPGA being capable of running simultaneous tasks, the relative delays between the starting times of eight-channel RF pulse sequences are shorter than those of the previous works, which is more beneficial to manipulate atoms at the given time. By pre-loading the control parameters of the DDS output signals into the internal register of AD9910 and making full use of the RAM mode, single tone mode, and DRG mode of DDSs, the program of AD9910 is completed in advance before the RF pulse sequence generation. Thus only modifying the states of the PROFILE [2:0] pins can make the output frequency and/or amplitude of RF signal be switched in the shorter time. The delay time of frequency switching, less than 119 ns, is quantitatively given by measuring the rise time of the light signal. Arbitrary RF pulse sequences, needed by the atom interferometers, can be completely generated by fast switching/sweeping the frequency, amplitude, and phase of RF pulses, and they have been applied to a dual-loop atom-interferometer gyroscope. This compact multi-channel generator with the fast switching/sweeping capability has beneficial applications for the integration of cold atom interferometers.

## V. ACKNOWLEDGEMENT

We acknowledge the financial support from the National Innovation Program for Quantum Science and Technology of China under Grant No. 2021ZD0300604, the National Natural Science Foundation of China under Grant Nos. 12104466, 11674362, 91536221, and 91736311, the Strategic Priority Research Program of Chinese Academy of Sciences under Grant No. XDB21010100, the Outstanding Youth Foundation of Hubei Province of China under Grant No. 2018CFA082, the Postdoctoral Innovation Research Posts in Hubei Province of China under Grant No. R20R0004, and the Youth Innovation Promotion Association of Chinese Academy of Sciences.